\documentclass[prl,aps,twocolumn,showpacs,nobibnotes,epsf]{revtex4}
\usepackage{graphicx}% Include figure files
\usepackage{dcolumn}% Align table columns on decimal point
\usepackage{bm}% bold math
\usepackage{SIunits}

\begin{document}
\title{ Transport properties and superconductivity in $Ba_{1-x}M_xFe_2As_2$ (M=La and K) with double FeAs layers }
\author{G. Wu, R. H. Liu, H. Chen, Y. J. Yan, T. Wu, Y. L. Xie, J. J. Ying, X. F. Wang, D. F. Fang}
\author{ X. H. Chen}
\altaffiliation{Corresponding author} \email{chenxh@ustc.edu.cn}
\affiliation{Hefei National Laboratory for Physical Science at
Microscale and Department of Physics, University of Science and
Technology of China, Hefei, Anhui 230026, P. R. China\\ }
\date{\today}

\begin{abstract}
We synthesized the samples $Ba_{1-x}M_xFe_2As_2$ (M=La and K) with
$ThCr_2Si_2$-type structure. These samples were systematically
characterized by resistivity, thermoelectic power (TEP) and Hall
coefficient ($R_H$). $BaFe_2As_2$ shows an anomaly in resistivity at
about 140 K. Substitution of La for Ba leads to a shift of the
anomaly to low temperature, but no superconducting transition is
observed. Potassium doping leads to suppression of the anomaly in
resistivity and induces superconductivity at 38 K as reported by
Rotter et al.\cite{rotter}. The Hall coefficient and TEP
measurements indicate that the TEP is negative for $BaFe_2As_2$ and
La-doped $BaFe_2As_2$, indicating n-type carrier; while potassium
doping leads to change of the sign in $R_H$ and TEP. It definitely
indicates p-type carrier in superconducting $Ba_{1-x}K_xFe_2As_2$
with double FeAs layers, being in contrast to the case of
$LnO_{1-x}F_xFeAs$ with single FeAs layer. A similar
superconductivity is also observed in the sample with nominal
composition $Ba_{1-x}K_xOFe_2As_2$.
\end{abstract}


\vskip 300 pt

\maketitle

Layered rare-earth metal oxypnictides LnOMPn (Ln=La, Pr, Ce, Sm;
M=Fe, Co, Ni, Ru and Pn=P and As) with ZrCuSiAs type
structure\cite{quebe,zimmer} have attracted great attention due to
the discovery of superconductivity at $T_c=26$ K in the iron-based
$LaO_{1-x}F_x$FeAs (x=0.05-0.12)\cite{yoichi}. Immediately, $T_c$
was drastically raised to 43 K in $SmO_{1-x}F_x$FeAs\cite{chenxh},
$T_c$=41 K in $CeO_{1-x}F_x$FeAs\cite{chen}, and 52 K in
$PrO_{1-x}F_x$FeAs\cite{ren}. These discoveries have generated much
interest for exploring novel high temperature superconductor, and
provided a new material base for studying the origin of high
temperature superconductivity. Such high-$T_c$ iron pnictides adopts
a layered structure of alternating Fe-As and Ln-O layers with eight
atoms in a tetragonal unit cell. Similar to the cuprates, the Fe-As
layer is thought to be responsible for superconductivity, and Ln-O
layer is carrier reservoir layer to provide electron carrier. In
order to induce the electron carrier, three different ways have been
used: (1) substitution of flourine for oxygen \cite{yoichi,chenxh};
(2) to produce oxygen deficiency\cite{ren1}; and (3) substitution of
$Th^{4+}$ for $Ln^{3+}$\cite{wang}. All these ways for inducing
electron carrier are limited in the carrier reservoir Ln-O layer by
substitution. The electron carrier induced transfers to Fe-As layer
to realize superconductivity except for the case Sr-doping in
LaOFeAs system\cite{wen}.

Very recently, the ternary iron arsenide $BaFe_2As_2$ shows
superconductivity at 38 K by hole doping with partial substitution
of potassium for barium\cite{rotter}. This material is
$ThCr_2Si_2$-type structure. There exists single FeAs layer in unit
cell in LnOFeAs system, while there are double FeAs layers in
$BaFe_2As_2$. The undoped material LaOFeAs shows an anomaly in
resistivity at 150 K which is associated with the structural
transition or SDW transition\cite{cruz,mcguire}. Susceptibility
measurements indicate that an antiferromagnetic phase transition
around 130 K occurs in $BaFe_2As_2$\cite{pfisterer}. Although Rotter
et al. claimed hole doping by substitution of potassium for barium
in superconductor $Ba_{1-x}K_xFe_2As_2$, no direct evidence supports
this speculation\cite{rotter}. Here we prepared single phase the
samples $Ba_{1-x}M_xFe_2As_2$ (M=La and K), and systematically
studied their transport properties (resistivity, Hall coefficient,
and thermoelectric power). It is found that $BaFe_2As_2$ shows an
anomaly in resistivity at about 140 K. Substitution of La for Ba
leads to a shift of the anomaly to low temperature, but no
superconducting transition is observed. Potassium doping leads to
suppression of the anomaly in resistivity and induces
superconductivity at 38 K. The Hall coefficient and TEP measurements
indicate that the TEP is negative for $BaFe_2As_2$ and La-doped
$BaFe_2As_2$, indicating n-type carrier; while potassium doping
leads to change of the sign in $R_H$ and TEP. It definitely
indicates p-type carrier in superconducting $Ba_{1-x}K_xFe_2As_2$
with double FeAs layers, being in contrast to the case of
$LnO_{1-x}F_xFeAs$ with single FeAs layer.

Polycrystalline samples of BaFe$_2$$As_2$,
Ba$_{1-x}$K$_x$Fe$_2$$As_2$ and Ba$_{1-x}$La$_x$Fe$_2$$As_2$ were
synthesized by solid state reaction method using BaAs, FeAs, Fe,
LaAs and K as starting materials. BaAs was pre-synthesized by
heating the mixture of Ba power and As power in an evacuated quartz
tube at 673 K for 4 hours. FeAs and LaAs were obtained by reacting
the mixture of element powers in evacuated quartz tubes at 873 K for
4 hours. The raw materials were accurately weighed according to the
stoichiometric ratio of BaFe$_2$$As_2$, Ba$_{1-x}$K$_x$Fe$_2$$As_2$
and Ba$_{1 -x}$La$_x$Fe$_2$$As_2$, then the weighed powders were
thoroughly grounded and pressed into pellets. The pellets were
wrapped with Ta foil and sealed in evacuate quartz tubes. The sealed
tubes were slowly heated to 1273 K and annealed for 20 hours. The
sample preparation process except for annealing was carried out in
glove box in which high pure argon atmosphere is filled. The sample
with nominal composition $Ba_{0.5}K_{0.5}OFe_2As_2$ was also
prepared in the method described above.

Figure 1 shows x-ray powder diffraction patterns for the samples:
$BaFe_2As_2$, $Ba_{0.85}La_{0.15}Fe_2As_2$,
$Ba_{0.6}K_{0.4}Fe_2As_2$ and $Ba_{0.5}K_{0.5}OFe_2As_2$. All
diffraction peaks in the patterns of $BaFe_2As_2$ and
\begin{figure}
\includegraphics[width=9cm]{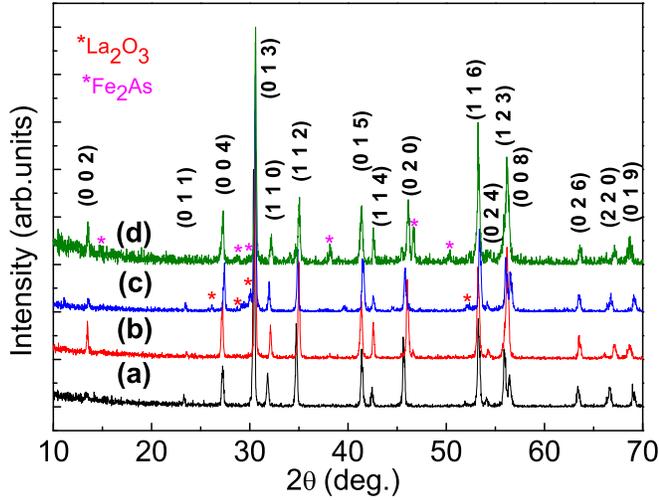}
\caption{(color online) X-ray powder diffraction patterns at room
temperature for the samples (a): $BaFe_2As_2$; (b):
$Ba_{0.6}K_{0.4}Fe_2As_2$; (c): $Ba_{0.85}La_{0.15}Fe_2As_2$; (d):
$Ba_{0.5}K_{0.5}OFe_2As_2$.
\\}
\end{figure}
$Ba_{0.6}K_{0.4}Fe_2As_2$ can be indexed by a tetragonal structure
with a=0.3961 nm  and c=1.3017 nm for the sample $BaFe_2As_2$,
a=0.3921 nm and c=1.3228 nm for the sample
$Ba_{0.6}K_{0.4}Fe_2As_2$, indicating that the samples is almost
single phase. It is found that K doping leads to an apparent
decrease in a-axis and an increase in c-axis. The lattice parameters
are consistent with the previous report\cite{rotter}. In order to
induce the electron into system, we tried to synthesize the sample
$Ba_{0.85}La_{0.15}Fe_2As_2$. Its x-ray powder diffraction pattern
is shown in Fig.1c, most of diffraction peaks can be indexed by
tetragonal structure with a=0.3957 nm and c=1.2989 nm, which is
slightly less than that of $BaFe_2As_2$. A trace of impurity phase
$La_2O_3$ is observed in the x-ray diffraction pattern, which could
arise from the reaction of unreacted La with oxygen in air. It
should be emphasized that the $ThCr_2Si_2$-type structure is also
formed with nominal composition $Ba_{1-x}K_xOFe_2As_2$ (x=0-0.5).
X-ray powder diffraction pattern is shown in Fig.1d. Except for the
diffraction peaks from impurity phase $Fe_2As$, all the peaks in the
pattern can be indexed to the $ThCr_2Si_2$-type structure with
a=0.3924 nm and c=1.3192 nm. This result seems to indicate that the
oxygen cannot be induced into the product. A similar case is also
found with nominal composition of $BaOFeAs$, $ThCr_2Si_2$-type
structure can also be formed with nominal composition $BaOFeAs$.

Figure 2 shows temperature dependence of resistivity for the samples
$BaFe_2As_2$, $Ba_{0.85}La_{0.15}Fe_2As_2$,
$Ba_{0.6}K_{0.4}Fe_2As_2$ and $Ba_{0.5}K_{0.5}OFe_2As_2$. The
resistivity of $BaFe_2As_2$ shows a linear temperature dependence
above a characteristic temperature of about 140 K, and show a
\begin{figure}
\includegraphics[width=9cm]{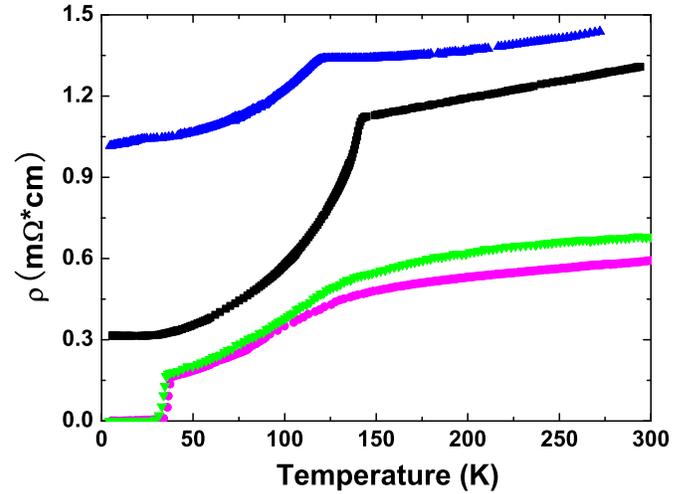}
\caption{(color online)  Temperature dependence of resistivity for
the samples (a): $BaFe_2As_2$ (squares); (b):
$Ba_{0.6}K_{0.4}Fe_2As_2$ (circles); (c):
$Ba_{0.85}La_{0.15}Fe_2As_2$ (up-triangles); (d):
$Ba_{0.5}K_{0.5}OFe_2As_2$ (down-triangles).
\\}
\end{figure}
steeply decrease with decreasing temperature below 140 K. These
results are consistent with previous report\cite{rotter}, and the
characteristic temperature is ascribed to antiferromagnetic phase
transition\cite{pfisterer}. As reported by Rotter et al., a
superconducting transition at about 38 K  is observed in resistivity
in K-doping sample $Ba_{0.6}K_{0.4}Fe_2As_2$. While no
superconducting transition down to 5 K is observed although the
characteristic temperature associated with magnetic transition
shifts to low temperature of $\sim 120$ K with La-doping in the
sample $Ba_{0.85}La_{0.15}Fe_2As_2$. It suggests that La-doping
suppresses the antiferromagnetic transition. We tried to increase
the La-doping to induce more electron into system, but found that
large amount of impurity phase shows up, so that La cannot induce to
the system by conventional solid state reaction. The sample
$Ba_{0.5}K_{0.5}OFe_2As_2$ shows the superconductivity at about 36
K. The temperature dependence of resistivity for the sample
$Ba_{0.5}K_{0.5}OFe_2As_2$ is similar to that observed in the sample
$Ba_{0.6}K_{0.4}Fe_2As_2$. It seems to indicate that the oxygen
cannot induce into the lattice since they show the same
superconductivity as that in $Ba_{1-x}K_xFe_2As_2$. It could be
amazing if the product did not include oxygen in conventional solid
state reaction with oxides as starting materials.

\begin{figure}
\includegraphics[width=9cm]{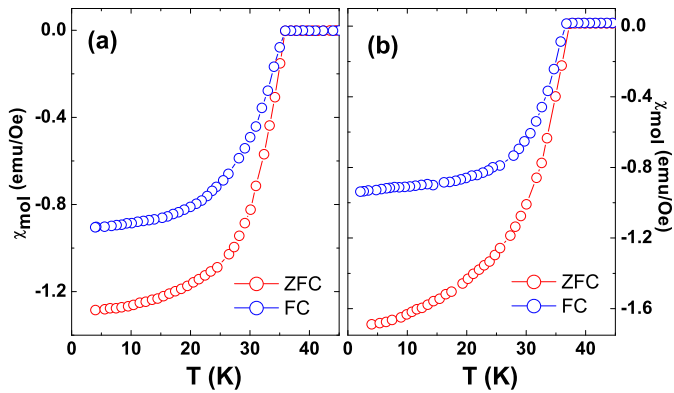}
\caption{(color online) Temperature dependence of susceptibility for
the samples (a): $Ba_{0.5}K_{0.5}OFe_2As_2$ and (b):
$Ba_{0.6}K_{0.4}Fe_2As_2$ under 10 Oe in zero-field cooled and field
cooled process.
\\}
\end{figure}

To confirm the superconductivity observed in resistivity for the
samples with nominal compositions $Ba_{0.6}K_{0.4}Fe_2As_2$ and
$Ba_{0.5}K_{0.5}OFe_2As_2$, the susceptibility measured under 10 Oe
in zero-field cooled (shielding) and field-cooled (Meissner) cycle
is shown in Fig.3. Fig.3 shows a superconducting transition for both
of the samples. The transition temperature is about 36 K for the
sample $Ba_{0.5}K_{0.5}OFe_2As_2$ and 38 K for the sample
$Ba_{0.6}K_{0.4}Fe_2As_2$, respectively. These transition
temperatures are consistent with that observed in resistivity. The
data in Fig.3 gives superconducting fraction of about $34\%$ and
about $18.7\%$ Meissner fraction at 5 K for sample
$Ba_{0.6}K_{0.4}Fe_2As_2$, and about $26\%$ shielding fraction and
about $18.3\%$ Meissner fraction for the sample
$Ba_{0.5}K_{0.5}OFe_2As_2$.

\begin{figure}
\includegraphics[width=9cm]{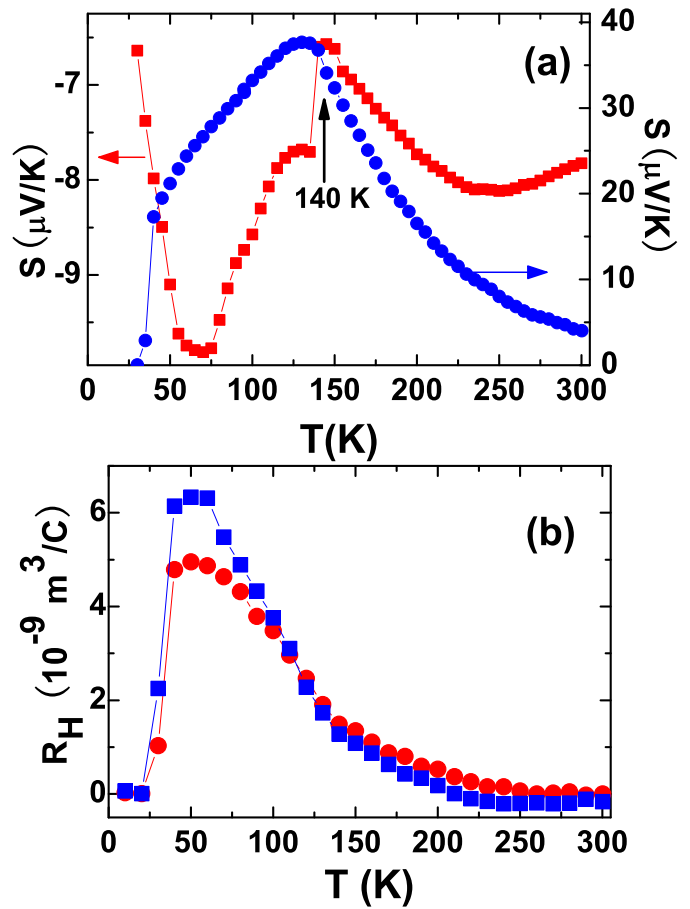}
\caption{(color online) Temperature dependence of (a):
thermoelectric power and (b): Hall coefficient. (a): for the samples
$BaFe_2As_2$ (squares),  $Ba_{0.6}K_{0.4}Fe_2As_2$ (circles); (b):
for the samples $Ba_{0.6}K_{0.4}Fe_2As_2$ and
$Ba_{0.5}K_{0.5}OFe_2As_2$.
\\}
\end{figure}

As claimed by Rotter et al.\cite{rotter}, hole doping is expected by
partial substitution of the barium with potassium in the samples
$Ba_{0.6}K_{0.4}Fe_2As_2$ and $Ba_{0.5}K_{0.5}OFe_2As_2$. In order
to provide the direct evidence to confirm it, the thermoelectric
power and Hall coefficient are systematically measured. Fig.4 shows
the temperature dependence of thermoelectric power and Hall
coefficient. TEP of $BaFe_2As_2$ is negative, being similar to the
parent compound LaOFeAs with single FeAs layer, but its magnitude is
less than that of parent compound LaOFeAs with single FeAs
layer\cite{mcguire}. TEP of $BaFe_2As_2$ shows a complicated
temperature dependence. TEP slightly increases with decreasing
temperature to about 240 K, and decreases with further decreasing.
At the characteristic temperature of 140 K, a big jump is observed
due to the antiferromagnetic transition or structural transition.
Below 140K, TEP continuously increases, while sharply decreases
below about 70 K. It should be addressed that all results discussed
here are well reproducible.  K-doping leads to change the sign of
TEP from negative to positive. It definitely indicates that the
K-doping induces the hole carrier into system. TEP monotonically
increases with decreasing temperature for the sample
$Ba_{0.6}K_{0.4}Fe_2As_2$. Below about 140 K, TEP decreases with
decreasing temperature down to about 40 K. At about 40 K, TEP
sharply decreases to zero, indicating a superconducting transition.
Figure 4b shows the temperature dependence of Hall coefficient for
the samples $Ba_{0.6}K_{0.4}Fe_2As_2$ and
$Ba_{0.5}K_{0.5}OFe_2As_2$. The $R_H$ of the two samples shows
similar temperature dependent behavior. $R_H$ is positive, being
consistent with that of TEP shown in Fig.4a, further indicating
hole-type superconductors $Ba_{1-x}K_{x}Fe_2As_2$ with double FeAs
layers. The $R_H$ is very small at room temperature, and negative
for the sample $Ba_{0.5}K_{0.5}OFe_2As_2$. With decreasing
temperature, the $R_H$ changes to positive and increases
monotonically down to the superconducting transition temperature.
These results provide direct evidence to confirm the p-type carrier
in $Ba_{1-x}K_{x}Fe_2As_2$ with double FeAs layers, being in
contrast to the case for $LnO_{1-x}F_xFeAs$ superconductors with
single FeAs layer. One open question is why n-type carrier can be
induced into the system with single FeAs layer, while p-type carrier
is induced to the superconductors with double FeAs layers.

In summary, the samples $Ba_{1-x}M_xFe_2As_2$ (M=La and K) with
$ThCr_2Si_2$-type structure were systematically characterized by
resistivity, thermoelectic power (TEP) and Hall coefficient ($R_H$).
Substitution of La for Ba leads to a shift of the anomaly to low
temperature, but no superconducting transition is observed.
Potassium doping leads to suppression of the anomaly in resistivity
and induces superconductivity at 38 K. The Hall coefficient and TEP
measurements indicate that the TEP is negative for $BaFe_2As_2$ and
La-doped $BaFe_2As_2$, indicating n-type carrier; while potassium
doping leads to change of the sign in $R_H$ and TEP. It definitely
indicates p-type carrier in superconducting $Ba_{1-x}K_xFe_2As_2$
with double FeAs layers, being in contrast to the case of
$LnO_{1-x}F_xFeAs$ with single FeAs layer. It is amazing that the
superconductivity can be also realized in the sample with nominal
composition $Ba_{0.5}K_{0.5}OFe_2As_2$. It deserves to check if the
oxygen was induced into lattice in the sample
$Ba_{0.5}K_{0.5}OFe_2As_2$.

\vspace*{2mm} {\bf Acknowledgment:} This work is supported by the
Natural Science Foundation of China and by the Ministry of Science
and Technology of China (973 project No: 2006CB601001) and by
National Basic Research Program of China (2006CB922005).


\begin{thebibliography}{00}

\baselineskip 9pt
\bibitem{rotter}
M. Rotter, M. Tegel and D. Johrendt, arXiv:0805.4630v1(2008).
\bibitem{zimmer}
B. I. Zimmer et al., \emph{J. Alloys and Compounds} {\bf 229},
238(1995).
\bibitem{quebe}
P. Quebe  et al.,  \emph{J. Alloys and Compounds} {\bf 302},
70(2000).
\bibitem{yoichi}
Y. Kamihara  et al., \emph{J. Am. Chem. Sco.} {\bf 130}, 3296(2008).
\bibitem{chenxh}
 X. H. Chen et al., Nature {\bf 354}, 761-762(2008).
\bibitem{chen}
G. F. Chen et al., Phys. Rev. Lett. (in press) (2008).
\bibitem{ren}
Z. A. Ren et al., arXiv:0803.4283v1(2008).
\bibitem{ren1}
Z. A. Ren et al., arXiv:0804.2582v1(2008)
\bibitem{wang}
C. Wang et al., arXiv:0804.4290v1(2008).
\bibitem{wen}
H. H. Wen, G. Mu, L. Fang, H. Yang, and X. Zhu, Europhys. Lett. {\bf
82}, 17009(2008).
\bibitem{cruz}
C. Cruz et al.,  arXiv:0804.0795(2008).
\bibitem{mcguire}
 M. A. McGuire et al., arXiv:0804.0796(2008).
\bibitem{pfisterer}
M. Pfisterer and G. Nagorsen, Z. Naturforsh. B: Chem. Sci. {\bf 38},
811(1983).


\end{thebibliography}
\end{document}